\documentclass[final]{svjour3}                     
\smartqed  
\usepackage{graphicx}
\journalname{Journal of Low Temperature Physics}

\begin{document}

\title{Breakdown of Potential Flow to Turbulence around a Sphere Oscillating in Superfluid $^4$He above the Critical Velocity. }


\titlerunning{Oscillating Sphere}        

\author{W. Schoepe \and R. H\"anninen \and
        M. Niemetz }


\institute{W. Schoepe \at Fakult\"at f\"ur Physik, Universit\"at Regensburg \\
              \email{wilfried.schoepe@ur.de} \and
R. H\"anninen \at Low Temperature Laboratory, O.V. Lounasmaa Laboratory, Aalto University, Finland \and
           M. Niemetz \at OTH Regensburg, Regensburg}

\date{}

\maketitle

\begin{abstract}

The onset of turbulent flow around an oscillating sphere in superfluid $^4$He is known to occur at a critical velocity $v_c \sim \sqrt{\kappa \,\omega }$ where $\kappa $ is the circulation quantum and $\omega $ is the oscillation frequency. But it is also well known that initially in a first up-sweep of the oscillation amplitude, $v_c$ can be considerably exceeded before the transition occurs, thus leading  to a strong hysteresis in the velocity sweeps. The velocity amplitude $v_c^* > v_c$ where the transition finally occurs is related to the density $L_0$ of the remanent vortices in the superfluid. Moreover, at temperatures below ca. 0.5 K and in a small interval of velocity amplitudes between $v_c$ and a velocity that is about 2\% larger, the flow pattern is found to be unstable, switching intermittently between potential flow and turbulence. From time series recorded at constant temperature and driving force the distribution of the excess velocities $\Delta v = v_c^* - v_c$ is obtained and from that the failure rate. Below 0.1 K we also can determine the distribution of the lifetimes of the phases of potential flow. Finally, the frequency dependence of these results is discussed. 

\end{abstract}

\section{Introduction}
\label{intro}

In this article the results of earlier work on the transition from potential flow to turbulence at velocity amplitudes above the critical velocity $v_c \sim \sqrt{\kappa \,\omega }$ (where $\kappa$ = 0.0997 mm$^2$/s is the circulation quantum and $\omega $ is the oscillation frequency) observed with a sphere oscillating in superfluid $^4$He are reviewed and extended. Our sphere has a radius of 0.1 mm and it is oscillating in a cell of cylindrical shape of 2 mm diameter and 1 mm height. The surface roughness of the sphere is estimated to be of the order of few micrometer and, therefore, the sphere is rough on the scale of the vortex core of 0.1 nm. At temperatures above 0.5 K a hysteretic behavior in the velocity amplitude vs. driving force amplitude was first reported in 1995 \cite {PRL1}. More recent data are shown in Fig.~1 \cite{JLTP2008}. What one generally observes is a linear increase of the velocity amplitude with increasing driving force. The linear drag force is due to thermally excited quasiparticles, i.e., ballistic phonons at low temperatures and normal phase drag (Stokes) in the hydrodynamic  regime above 1 K. At larger drive a strong nonlinear drag is observed when turbulence (vortices) are created. There is a hysteresis: only when reducing the drive in the turbulent regime the critical velocity $v_c$ can be determined, because in the initial up-sweep potential flow breaks down only at velocities $v_c^*$ larger than $v_c$.  

\begin{figure}[t]
\centerline{\includegraphics[width=0.60\textwidth]{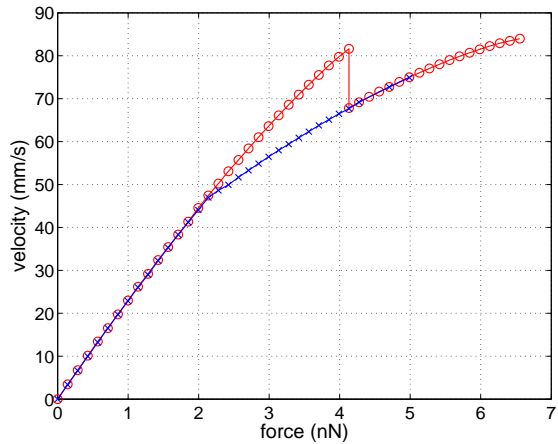}}
\caption{(From Ref.2) (Color online) Velocity amplitude as a function of the driving force for a 100$\mu$m sphere oscillating at 236 Hz in superfluid $^4$He at 1.90 K. Red circles: data taken with increasing drive; blue x :  decreasing drive. The initial breakdown of the potential flow occurs at $v_c^*$ = 82 mm/s. Because of the strong hysteresis $v_c$ = 46 mm/s can only be determined with the down sweep.}
\label{fig:1}
\end{figure}

\begin{figure}[ht]
\centerline{\includegraphics[width=0.8\linewidth, keepaspectratio]{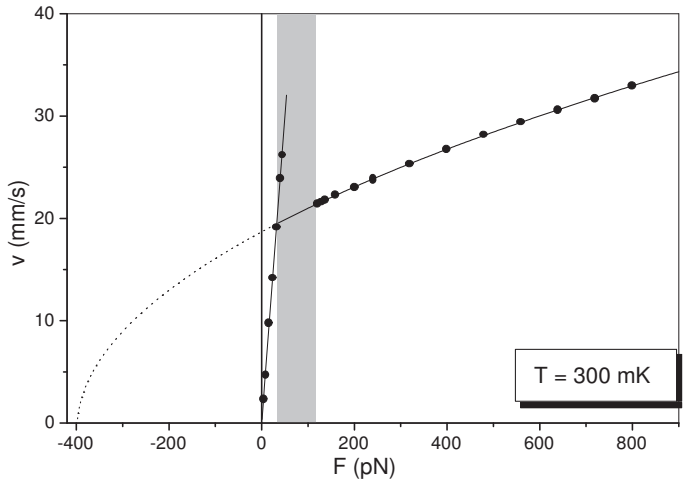}}
\caption{(From Ref.5) Velocity amplitude $v$ of the oscillating sphere (frequency 114 Hz) as a function of the driving 
force $F$ at $300\,\mbox{mK}$. At small drives we have potential flow where the linear drag is determined by ballistic phonon scattering. The shaded area above the critical velocity $v_c \approx $ 20 mm/s indicates the unstable regime where the flow switches intermittently between potential flow and turbulence (as shown in Fig.~3). At larger driving forces we observe apparently stable nonlinear turbulent drag where $F(v) \propto (v^2 - v_c^2)$ (when the small laminar drag is subtracted).}
\label{fig:2}       
\end{figure}
In a set of careful experiments with a vibrating wire the Osaka group \cite {Yano} showed in 2007 that no transition to turbulence occurred up to very large velocity amplitudes of ca. 1 m/s (corresponding to oscillation amplitudes of 100 $\mu$m) if the superfluid was prepared in a state without any remanent vortices. Obviously, some initial vortices must exist in the fluid within the range of the oscillation amplitude for the vibrating object to produce turbulence.

\section{Intermittent switching between potential flow and turbulence}

\subsection{Temperatures between 0.5 K and 0.1 K}

The hysteresis which can be observed almost up to the Lambda transition changes into an instability of the flow pattern below ca. 0.5 K where intermittent switching between potential flow and turbulent flow around an oscillating sphere was observed at velocity amplitudes slightly above $v_c$, see Fig.~2 and Fig.~3. (Experimental details are described in \cite {Kerscher,Niemetz}.)

\begin{figure}[ht]
\centerline{\includegraphics[width=0.8\textwidth]{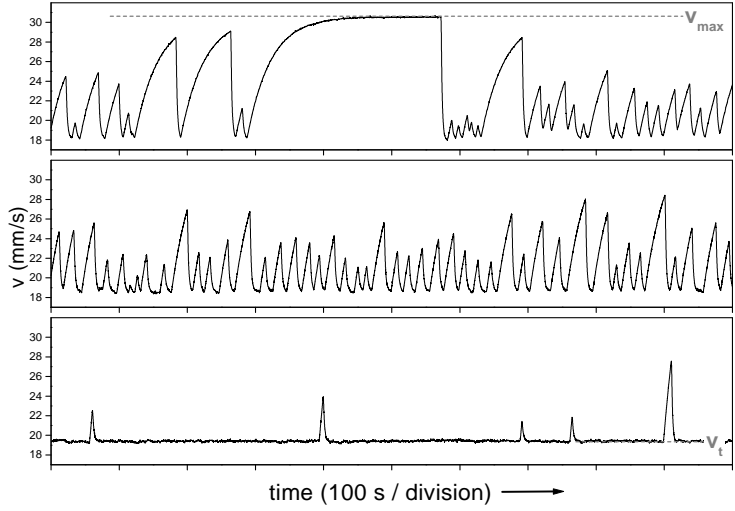}}
\caption{(From Ref.5) Three time series of the velocity amplitude at 300 mK and 114 Hz at three different driving forces (in pN): 47, 55, 75 (from top to bottom). The low level $v_t$ corresponds to turbulent drag while the increase of the velocity amplitude occurs during potential flow, occasionally reaching the maximum value $v_{max}$ given by ballistic phonon drag.} 
\label{fig:3}       
\end{figure}

During a turbulent phase the drag is given by the large turbulent drag force and, hence, the velocity amplitude is low. (The statistical properties of the turbulent phases were analyzed recently in some detail, see \cite{JLTP173,arXiv}). When turbulence breaks down the velocity amplitude begins to grow due to the much smaller phonon drag. Because the velocity amplitudes are above $v_c$ these phases of potential flow break down at some velocity $v_c^* > v_c$ and a new turbulent phase is observed. The distribution of the maximum of the excess velocities $\Delta v = v_c^* - v_c$ above the turbulent level is analyzed. This is done by evaluating the probability $R(\Delta v)$ that a given $\Delta v$ is exceeded (``reliability function"), see Fig.~4a. From a fit to the data we find that $R(\Delta v) = \exp(-(\Delta v / v_w)^2)$ is a Rayleigh distribution and the fitting parameter $v_w$ is independent of temperature and driving force, see Fig.~4b. It is a property of the Rayleigh distribution that $v_w$ is the rms value of the distribution.

\begin{figure}
\begin{center} 
 \includegraphics[width=0.8\textwidth]{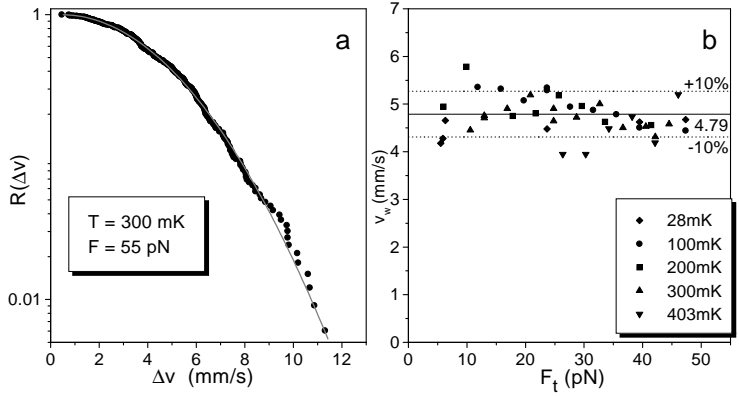}
\end{center}
\caption{(From Ref.5) Statistical properties of the phases of potential flow shown in Fig.~3. a) The reliability function $R(\Delta v) = \exp(-(\Delta v / v_w)^2)$ is a Rayleigh distribution. b) The fitting parameter $v_w$ is independent of temperature and driving force (phonon drag subtracted).}
\label{fig:4}       
\end{figure}

From the Rayleigh distribution we obtain the failure rate $\Lambda (\Delta v)$ which is the conditional probability that the potential flow decays in a small interval just above $\Delta v$, provided it has not decayed until $\Delta v$:

\begin{equation}
\Lambda(\Delta v)=-\frac{d\ln R}{d\Delta v}= \frac{2\Delta v}{ v^2_w}.\label{1}
\end{equation}

We note that the failure rate is proportional to $\Delta v$ and, hence, to the increase of the oscillation amplitude $\Delta a = \Delta v / \omega$. This result will be of significance when we consider the effect of the remanent vortex density $L_0$ on the critical velocity, see Section 4.

Another general property of this distribution can be found when changing the variable from $\Delta v$ to the lifetime $t$ of the potential flow:

\begin{equation}
\Lambda(t)=\Lambda(\Delta v(t)) \cdot \frac{d\Delta v}{dt} = \frac{2 \Delta v(t)}{v_w^2}  \cdot \frac{d\Delta v}{dt}=\frac{1}{v^2_w}\frac{d(\Delta v)^2}{dt},\label{2} 
\end{equation} 
where in our case $\Delta v(t)$ is due to the relaxation to the maximum amplitude $v_{max}$ given by ballistic phonon scattering (see upper trace in Fig.~3):
\begin{equation}
\Delta v(t) = \Delta v_{max} (1 - \exp{(-t/\tau))},
\end{equation}
where $\Delta v_{max}$ = $v_{max} - v_c$ and $\tau$ = 2m/$\lambda$ is the time constant, $\lambda$ is the friction coefficient of phonon scattering and m is the mass of the sphere (27 $\mu$g).
Initially we have a linear increase of $\Delta v(t)$ and therefore $\Lambda(0) = 0$. At a time $t$ = $\tau \ln{2}$ the failure rate has a maximum. And, if the maximum level $\Delta v_{max}$ is approached for large $t$, we find $\Lambda  \rightarrow 0$. Thus the flow becomes stable although the velocity $v_{max}$ is clearly above $v_c$. Nevertheless, these metastable states of potential flow ultimately will also break down, in our experiments  after a mean lifetime of 25 min, because natural radioactivity occasionally produces vorticity in the fluid, see \cite{Kerscher,Niemetz}.

\subsection{Temperatures below 0.1 K}
Below 100 mK the phonon drag is negligibly small, only the residual damping of the empty cell $\lambda_{res}$ = 4.4 $10^{-11}$ kg/s \cite{Niemetz} determines the oscillation amplitude. Equivalent is a time constant $\tau$ = 2m/$\lambda_{res}$ = 1200 s. Therefore, the exponential function in (3) can be expanded:
\begin{figure}[!b]
\centerline{\includegraphics[width=0.8\textwidth,clip=true]{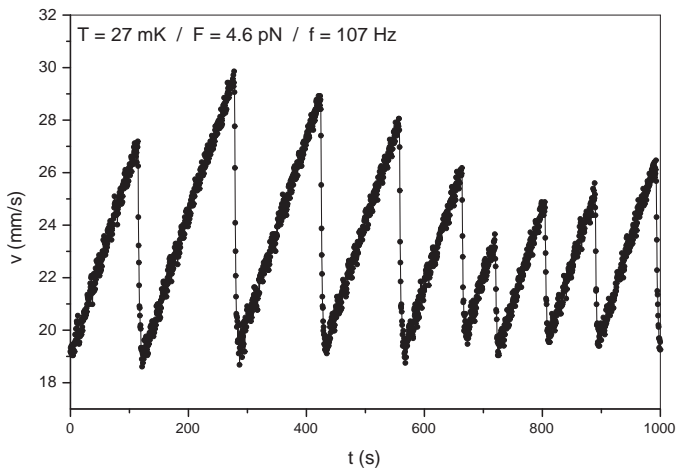}}
\caption{Time series of the breakdown of the phases of potential flow recorded at 27 mK with a driving force of 4.6 pN. The linear increase of $v(t)$ during potential flow is described by Eq.4, see text. The lifetimes of the turbulent phases cannot be resolved (not even on an expanded time scale).}
\label{fig:5}       
\end{figure}\\

\begin{equation}
\Delta v(t) = \left(\frac{F}{\lambda_{res}} - v_c\right) \frac{t}{\tau} = \left(\frac{F}{2m} - \frac{v_c\lambda_{res}}{2m}\right)t.
\end{equation}
In Fig.~5 the time series recorded at 27 mK shows a stochastic saw tooth pattern. The slope of the linear increase of the velocity amplitude during potential flow amounts to 6.9 $10^{-5}$ m/s$^2$. Inserting in (4) the relevant numbers for $F$ = 4.6 pN, $v_c$ = 19 mm/s, $m$ = 27 $\mu$g, and $\lambda_{res}$ = 4.4 $10^{-11}$ kg/s, we find a slope of  7.0 $10^{-5}$ m/s$^2$, in agreement with the experimental value. From (4) follows that the lifetimes of the phases of potential flow now also have a Rayleigh distribution. From the measured rms value $v_w$ = 6.5 mm/s of the $\Delta v$ distribution we obtain the rms lifetime $t_w = 2 m v_w/ (F - v_c\lambda_{res})$ = 93 s. 

The second term in the brackets of (4) which is due to the residual damping, is here 18\% of the first one and decreases further at larger drives. Ultimately, only the first term will dominate, i.e., we may neglect damping and approach the case of an undamped oscillator:  
\begin{equation}
\Delta v(t) = \frac{Ft}{2m}.
\end{equation}
The rms lifetime $t_w$ of the potential flow is then proportional to $1/F$:
\begin{equation}
t_w = 2 m v_w/ F ,
\end{equation}
and the failure rate (2) would simply be given by
\begin{equation}
\Lambda(t) = \left(\frac{F}{2 m v_w}\right )^2 2\,t = \frac{2\,t}{t_w^2}.
\end{equation}

\section{The frequency dependence of $v_w$}
\begin{figure}[b]
\centerline{\includegraphics[width=0.8\textwidth,clip=true]{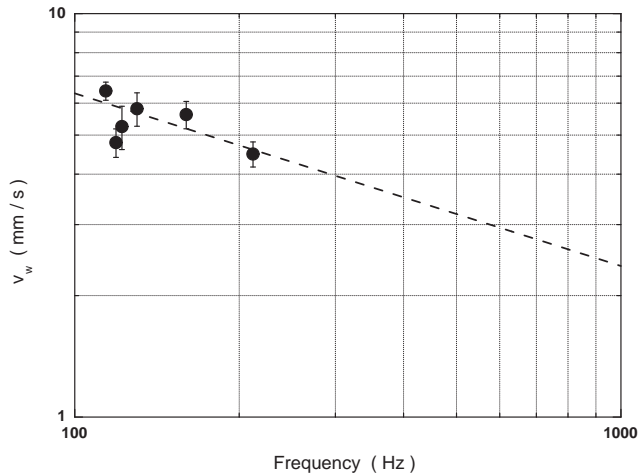}}
\caption{The rms excess velocities $v_w$ for six different frequencies. Each data point is obtained from the average of several individual time series recorded at different temperatures and driving forces. The error bars represent the standard deviations. A power law fit that takes into account the error bars indicates a slope of $-0.43 \pm 0.14$.}
\label{fig:6}       
\end{figure}
We observe a weak frequency dependence of $v_w$, see Fig.~6. A power law fit to the data indicates a slope of $- 0.43 \pm 0.14$. This may be interpreted as an $\omega ^{-1/2}$ law. However, our frequency range is rather limited ($\sim$ factor 2) and there is some scatter. Therefore, this dependence is not firmly established. On the other hand, the data point at 212 Hz is the mean of 14 individual $v_w$ data measured at different temperatures (28 mK and 300 mK) and different driving forces and, therefore, cannot simply be neglected. A theoretical interpretation of this result is not available as long as we do not have a theory of the transition to turbulence in oscillating superflows. However, a strong support of this frequency dependence of $v_w$ comes from the failure rate (7). In that case it follows that $\Lambda (t) \propto \omega \, t$, i.e., the failure rate is proportional to the number of the completed cycles. This is what one would naturally assume.

\section{Discussion}
The analysis of our data allows us to obtain informations concerning the properties of the instable phases of potential flow above the critical velocity. The failure rate (1) is proportional to the increase of the oscillation amplitude beyond $v_c$, and it is independent of temperature and driving force. The final breakdown of potential flow at $v_c^* > v_c$ is known to be affected by the remanent vorticity and occurs statistically during the switching process at mK temperatures. Moreover, the stability of the phases of potential flow when $v_{max}$ is reached (see upper trace in Fig.~3) leads us to assume that at breakdown the critical oscillation amplitude is determined by the intervortex spacing $l_0 = L_0^{-1/2}$ of the remanent vorticity, which in this case cannot be reached by the sphere: $l_0 > v_{max}/\omega \approx$ 40 $\mu$m that is less than half of the radius of the sphere. Introducing also at $v_c$ the intervortex spacing $l_c = L_c^{-1/2}$ we assume, without having a rigorous theory, that the critical oscillation amplitudes are proportional to the intervortex spacings:  

\begin{equation}
\frac{l_c}{l_0} = \frac{v_c}{v_c^*}.
\end{equation}
From the hysteresis shown in Fig.~1 we find from the velocities $v_c^*$ = 82 mm/s  and $v_c$ = 46 mm/s a ratio of 0.56 for the spacings and hence a ratio $L_0 / L_c$ = 0.31.
Defining $\Delta l = l_0 - l_c $ we obtain from (8) the relation

\begin{equation}
\frac{\Delta v}{v_c} = \frac{\Delta l}{l_c}.
\end{equation}
The rms value $v_w$ of the $\Delta v$ distribution is now related to the rms value $l_w$ of the $\Delta l$ distribution:

\begin{equation}
\frac{v_w}{v_c} = \frac{l_w}{l_c}.
\end{equation}
Inserting $v_w$ = 6.5 mm/s and $v_c$ = 19 mm/s (see Figs.~3 and 5) we get $l_w/l_c$ = 0.34 and thus an average $l_0$ = 1.34 $l_c$ or an average $L_0/L_c$ = 0.56. This means that on the average 56 \% of the vortex density at $v_c$ have survived the breakdown of the turbulent phase. But as long as $L_c$ is unknown for oscillatory flow we cannot determine absolute values of the vortex densities. Also, since our sphere radius may be larger than the intervortex distance it is rather peculiar that the transition to turbulence is not triggered by the vortices likely to be attached to the sphere. However, it is possible that such vortices are absent, and in the laminar state the surface of the sphere becomes free from vorticity or is only covered by very tiny loops which require a large velocity in order to expand. The transition to turbulence can then occur at oscillation amplitudes that are smaller than the radius of the sphere, see above. This would happen when the sphere becomes again in contact with the tangle that still remains in the vicinity of the sphere. The correct physical picture requires experiments that could visualize the vortices around the sphere or more realistic computer simulations.

Hysteresis and switching of the flow have been observed also with vibrating tuning forks and wires \cite{george1,yano2005,yano2007} and the significance of remanent vorticity for the critical velocity $v_c^*$ at breakdown was demonstrated \cite{Yano,yano2010}. In our work we are relating $v_c^*$ directly to the intervortex spacing of the remanent vortex density. 

Discussing the distribution of the lifetimes of the potential flow we note that because of the nonlinear dependence of $\Delta v (t)$ in (3) the distribution is very complicated. However, the situation becomes much simpler at low temperatures, see Section 2.2, where $\Delta v$ is proportional to the lifetime, see (4). The lifetimes are now also following a Rayleigh distribution and the rms lifetime decreases with increasing driving force. This may be compared with the recent experimental result of the Lancaster group on intermittent switching of the flow of around a vibrating tuning fork at mK temperatures \cite{george2}. These authors found that the average lifetimes of potential flow decrease towards larger flow velocities, in qualitative agreement with our rms value $t_w$ (6).

In summary, our data can be analyzed even in more detail and without the assumption (8) when a theory of the transition to turbulence in oscillatory superflows will be available.

\begin{acknowledgements}
We are grateful to Jan J\"ager and Hubert Kerscher for their co-operation. W.S. acknowledges discussions with Matti Krusius (Aalto University, Finland) and Shaun Fisher (Lancaster University, UK). R.H is supported by the Academy of Finland. 
\end{acknowledgements}

\end{document}